\documentclass[12pt]{article}

\usepackage{amsmath,amssymb}
\usepackage[usenames,dvipsnames,svgnames,table]{xcolor}
\usepackage[colorlinks=true,citecolor=blue]{hyperref}
\usepackage{natbib}
\usepackage{graphicx}
\usepackage{setspace}
\usepackage{framed}

\begin{document}

\title{Molecular Clock Dating using MrBayes}
\author{Chi Zhang$^{1,2}$}
\maketitle

\noindent{\it
$^1$Department of Bioinformatics and Genetics, Swedish Museum of Natural History, 11418 Stockholm, Sweden; \\
$^2$Key Laboratory of Vertebrate Evolution and Human Origins of Chinese Academy of Sciences, Institute of Vertebrate Paleontology and Paleoanthropology, Chinese Academy of Sciences, Beijing 100044, China; \\
\newline E-mail: zhangchi@ivpp.ac.cn
}

\newpage
\section*{Abstract}

This paper provides an overview and a tutorial of molecular clock dating using MrBayes, which is a software for Bayesian inference of phylogeny. 
Two modern approaches, total-evidence dating and node dating, are demonstrated using a dataset of Hymenoptera with molecular sequences and morphological characters. 
The similarity and difference of the two methods are compared and discussed. 
Besides, a non-clock analysis is performed on the same dataset to compare with the molecular clock dating analyses.

\bigskip
\noindent {\it Keywords}: Bayesian phylogenetic inference, molecular clock dating, MCMC, MrBayes

\section{Introduction}

MrBayes is a software for Bayesian phylogenetic inference \citep{Huelsenbeck:2001jt, Ronquist:2003gt} and has become widely used by biologists \citep{VanNoorden:2014hn}.
Many new features have been implemented since version 3.2 \citep{Ronquist:2012ir}, 
including species tree inference under the multi-species coalescent model \citep{Liu:2008hl, Liu:2009hf};
compound Dirichlet priors for branch lengths \citep{Rannala:2012ke, Zhang:2012ke};
marginal model likelihood estimation using stepping-stone sampling \citep{Xie:2011it};
topology convergence diagnostics using the average standard deviation of split frequencies \citep{Lakner:2008dn};
BEAGLE library support \citep{Ayres:2012gf} and parallel computing using MPI \citep{Altekar:2004jz}.

In addition to these features, the focus of this study is its functionality of divergence time estimation using node dating \citep[e.g.,][]{Yang:2006eu, Ho:2009gn} and total-evidence dating \citep{Ronquist:2012ea, Zhang:2016kf} approaches under relaxed molecular clock models \citep{Huelsenbeck:2000ic, Thorne:2002il, Drummond:2006kt, Lepage:2007bq}.
Both dating techniques are implemented in the Bayesian framework, such that the diversification process and the fossil information are incorporated in the priors.
However, they do have significant differences.
In node dating, only molecular sequences from extant taxa are used, and one or more internal nodes of the extant taxa tree are calibrated by user-specified prior distributions, typically derived from the fossil record, to estimate the ages of all other nodes in the tree.
In total-evidence dating, extant and extinct taxa are all included in the tree, and morphological characters are coded for fossil and extant taxa in a combined matrix with molecular sequences.
The age of each fossil is assigned a prior distribution directly.

Several steps are involved in a Bayesian molecular clock dating analysis, importantly including partitioning the data, specifying evolutionary models, calibrating internal nodes or fossils, and setting priors for the tree and the other parameters.
These will be demonstrated in the Tutorial section.

\section{Hymenoptera data}

The original data analyzed by \citet{Ronquist:2012ea, Zhang:2016kf} includes 60 extant and 45 fossil Hymenoptera (wasps, ants, and bees) and eight outgroup taxa.
The data were divided into eight partitions as follows: (1) morphology, (2) 12S and 16S, (3) 18S, (4) 28S, (5) 1\textsuperscript{st} and 2\textsuperscript{nd} codon positions of CO1, (6) 3\textsuperscript{rd} codon positions of CO1 (but not used in the analyses), (7) 1\textsuperscript{st} and 2\textsuperscript{nd} codon positions of both copies of EF1$\alpha$, and (8) 3\textsuperscript{rd}  codon positions of both copies of EF1$\alpha$.

The full data takes days to run. For illustration purpose, the data is truncated into ten extant taxa (nine Hymenoptera and one Raphidioptera) and ten fossils, with 200 morphological characters, 100 sites from 16S, and 210 sites from EF1$\alpha$.
In the tutorial below, this dataset is analyzed by both total-evidence dating and node dating approaches under diversified sampling of extant taxa \citep{Hohna:2011hd, Zhang:2016kf}, followed by a non-clock analysis under the compound Dirichlet prior for branch lengths \citep{Rannala:2012ke, Zhang:2012ke}.

\section{Tutorial}

\subsection{Getting started}

The program MrBayes is available from \url{https://github.com/NBISweden/MrBayes}, including pre-compiled executables for macOS and Windows, and source code for all platforms. The current version is 3.2.7, which is used here.
The truncated dataset {\tt hym.nex} is in the {\tt doc/tutorial} folder within the release.
For convenience, it is recommended to put it in the same directory as the executable (e.g., named {\tt mb} in macOS/Linux or {\tt mb.exe} in Windows).
The file {\tt hym.nex} in the NEXUS format can be opened by a text editor.
The data matrix is at the beginning, including morphological characters and molecular sequences.
Following the data block, users can write MrBayes commands within the mrbayes block (each ends with a semicolon).
These commands will be executed automatically when the data is read in.
The text within a pair of square brackets are comments and will be ignored by the program.

In terminal (macOS/Linux) or command prompt (Windows), navigate to the folder containing the executable and data file using the {\tt cd} command, and launch MrBayes using {\tt ./mb} (macOS/Linux) or {\tt mb.exe} (Windows). The following header will appear.
\begin{framed}
{\tt \center   MrBayes 3.2.7 x86\_64                \\
           (Bayesian Analysis of Phylogeny)         \\
   Distributed under the GNU General Public License \\
MrBayes > }
\end{framed}
\noindent The prompt at the bottom means that MrBayes is running and ready for your commands. 
Simply use
\begin{framed}
{\tt \noindent
execute hym.nex}	
\end{framed}
\noindent to read in the data.

\subsection{Data partitions}

When the data is read in, the commands for defining the partitions are also executed.
\begin{framed}
{\tt \noindent
charset MV = 1-200      \\
charset 16S = 201-300   \\
charset Ef1a = 301-510  \\
charset Ef1a12 = 301-510\textbackslash 3 302-510\textbackslash 3 \\
charset Ef1a3 = 303-510\textbackslash 3         \\
partition four = 4: MV, 16S, Ef1a12, Ef1a3      \\
set partition = four}
\end{framed}
\noindent Here we define four partitions:
the morphological, 16S, 1\textsuperscript{st} and 2\textsuperscript{nd} codon positions of Ef1$\alpha$, and 3\textsuperscript{rd} codon positions of Ef1$\alpha$.

\subsection{Substitution models}

For the morphological partition, the Mkv Model \citep{Lewis:2001wu} is used with variable ascertainment bias (only variable characters scored), equal state frequencies and gamma rate variation across characters.
The constant characters are thus excluded.
\begin{framed}
{\tt \noindent
exclude 7 31 61 83 107 121 122 133 182 183 198 \\
lset applyto = (1) coding = variable rates = gamma}
\end{framed}
\noindent If instant change is only allowed between adjacent states (e.g., 0 $\leftrightarrow$ 1 and 1 $\leftrightarrow$ 2 but not 0 $\leftrightarrow$ 2), these characters are specified using 
\begin{framed}
{\tt \noindent
ctype ordered: 20 23 27 30 36 41 42 44 46 48 59 65 75 78 79 89 99 112 117 134 146 157 159 171 185 191 193 196}
\end{framed}
\noindent The other characters are thus allowed to change instantly from one state to another.

For the molecular partitions, the general time-reversible model is used with gamma rate variation across sites (GTR$+\Gamma$) \citep{Yang:1994vo, Yang:1994wd}.
\begin{framed}
{\tt \noindent
lset applyto = (2,3,4) nst = 6 rates = gamma}
\end{framed}
\noindent The prior for the gamma shape parameter is exponential(1.0), which can be changed using {\tt prset shapepr}. We keep the default here.

Different partitions are assumed to have independent substitution parameters, thus we unlink them.
The partition-specific rate multipliers are used to account for rate variation across partitions with average to 1.0.
\begin{framed}
{\tt \noindent
unlink statefreq = (all) revmat = (all) shape = (all) \\
prset applyto = (all) ratepr = variable}
\end{framed}

\subsection{Relaxed clock models}

The relaxed clock models account for evolutionary rate variation over time and among branches, and it is now standard practice to accommodate such variation in dating analyses.
There are three relaxed clock models implemented in MrBayes: compound Poisson process \citep[CPP,][]{Huelsenbeck:2000ic}, autocorrelated lognormal \citep[TK02,][]{Thorne:2002il} and independent gamma rate \citep[IGR,][]{Lepage:2007bq}.
However the CPP model is computationally not compatible with total-evidence dating, thus we only focus on the IGR and TK02 models.

\begin{figure}[h]
\center
\includegraphics[width=0.6\textwidth]{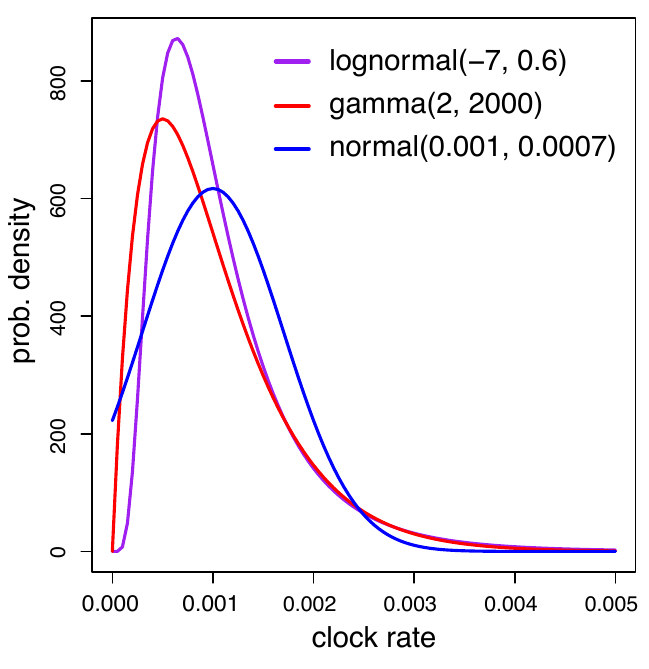}
\caption{Probability density functions of normal, lognormal and gamma distributions, all with mean 0.001 and standard deviation 0.0007.}
\label{Fig_ClockRatePr}
\end{figure}

The mean clock rate (mean substitution rate per site per Myr) is assigned a lognormal(-7,0.6) prior, with mean $e^{(-7 + 0.6^2/2)} = 0.001$ and standard deviation $\sqrt{(e^{0.6^2} -1) e^{(-2 \times 7 + 0.6^2)}} = 0.0007$.
\begin{framed}
{\tt \noindent
prset clockratepr = lognorm(-7,0.6)}
\end{framed}
\noindent There are several options for the clock rate prior, including {\tt fixed}, {\tt normal} (truncated at zero), {\tt lognormal}, and {\tt gamma}.
The probability density functions (all with mean 0.001 and standard deviation 0.0007) are shown in Figure \ref{Fig_ClockRatePr}.

The relative clock rates, which are multiplied by the mean clock rate, vary differently along the branches of the tree in different models.
The TK02 model assumes that the relative rate changes along the branches as Brownian motion on the log scale, starting from 1.0 (0.0 on the log scale) at the root.
The rate at the end of a branch is lognormal distributed with mean equal to the rate at the beginning of the branch.
Thus the rates at different branches are autocorrelated.
\begin{framed}
{\tt \noindent
prset clockvarpr = tk02 \\
prset tk02varpr = exp(1)}
\end{framed}
\noindent The IGR model assumes that the relative rate at each branch follows an independent gamma distribution with mean 1.0.
The variance is proportional to the branch length.
\begin{framed}
{\tt \noindent
prset clockvarpr = igr \\
prset igrvarpr = exp(10)}
\end{framed}
\noindent Note that when the relative rates are all fixed to 1.0 in the TK02 or IGR model, it becomes the strict clock model.

Before the total-evidence dating and node dating analyses, we first define the outgroup, fossil taxa, and some constraints for later use. Note these constraints are {\it not} enforced until we set {\tt topologypr} explicitly (see below).
\begin{framed}
{\tt \noindent
outgroup Raphidioptera  \\
taxset fossils = Asioxyela Nigrimonticola Xyelotoma Undatoma Dahuratoma Cleistogaster Ghilarella Mesorussus Prosyntexis Pseudoxyelocerus \\
constraint HymenFossil = 2-.    \\
constraint Hymenoptera = 2-10   \\
constraint Holometabola = 1-10  \\
constraint Tenthredinidae = 3-5 \\
constraint CepSirOruApo = 7-10}
\end{framed}

\subsection{Total-evidence dating}

In the following, we assign priors for the fossils from the geological times.
This is a typical step in total-evidence dating, where we calibrate the fossil taxa instead of the internal nodes of the tree.
\begin{framed}
{\tt \noindent
calibrate  \\
\indent Asioxyela = unif(228,242)      \\
\indent Nigrimonticola = unif(152,163) \\
\indent Xyelotoma = unif(152,163)      \\
\indent Undatoma = unif(145,152)       \\
\indent Dahuratoma = fixed(134)        \\
\indent Cleistogaster = unif(168,191)  \\
\indent Ghilarella = unif(113,125)     \\
\indent Mesorussus = unif(94,100)      \\
\indent Prosyntexis = unif(80,86)      \\
\indent Pseudoxyelocerus = fixed(182)  \\
prset nodeagepr = calibrated}
\end{framed}
\noindent The last command {\tt nodeagepr} is {\it necessary} to enable the calibrations.

\begin{figure}[h]
\center
\includegraphics[width=0.9\textwidth]{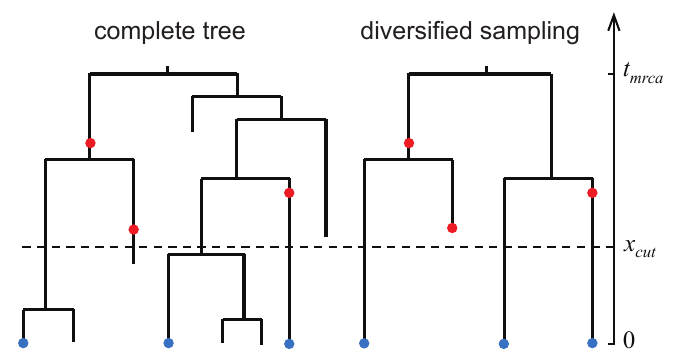}
\caption{The fossilized birth-death (FBD) process and diversified sampling of extant taxa. Exactly one representative taxa per clade descending from time $x_{cut}$ is sampled (blue dots). The fossils are sampled with a constant rate between $t_{mrca}$ and $x_{cut}$ (red dots).}
\label{Fig_FBDTree}
\end{figure}

The speciation, extinction, fossilization, and sampling process is explicitly modeled using the fossilized birth-death (FBD) process \citep{Stadler:2010fn, Heath:2014hn, Gavryushkina:2014fw, Zhang:2016kf}.
Diversified sampling \citep{Zhang:2016kf} is arguably suitable for such higher-level taxa, which assumes exactly one representative extant taxa per clade descending from time $x_{cut}$ is sampled, and the fossils are sampled with a non-zero rate before $x_{cut}$ and zero after.
A fossil can be either a tip or a sampled ancestor (Fig. \ref{Fig_FBDTree}).

The model has four parameters: speciation rate $\lambda$, extinction rate $\mu$, fossil discovery rate $\psi$, and extant sample proportion $\rho$.
For inference, we reparametrize the parameters as $d = \lambda - \mu$, $r = \mu / \lambda$, and $s = \psi/(\mu + \psi)$, so that the latter two parameters range from 0 to 1.
$\rho$ is fixed to 0.0001, based on the living number of Hymenoptera at about $10/0.0001=100,000$.
\begin{framed}
{\tt \noindent
prset brlenspr = clock:fossilization \\
prset samplestrat = diversity        \\
prset sampleprob = 0.0001            \\
prset speciationpr = exp(10)         \\
prset extinctionpr = beta(1,1)       \\
prset fossilizationpr = beta(1,1)    \\
prset treeagepr = offsetexp(300,390)}
\end{framed}
 
\begin{figure}[h]
\center
\includegraphics[width=0.6\textwidth]{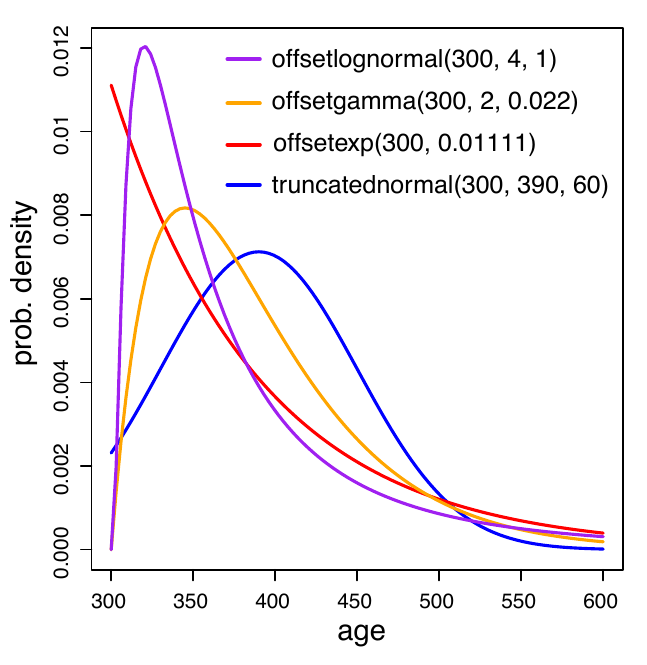}
\caption{Probability density functions of offset lognormal, offset gamma, offset exponential, and truncated normal distributions, all with mean 390 and minimum 300.}
\label{Fig_TreeAgePr}
\end{figure}

The FBD prior is conditioned on the root age ($t_{mrca}$), so it is important to set it properly.
Here we use an offset exponential distribution with minimal age 300 Ma and mean age 390 Ma.
Several available probability densities all with mean 390 and minimal 300 are shown in Figure \ref{Fig_TreeAgePr}.

An alternative tree prior that can be used in the total-evidence dating approach is the uniform prior ({\tt clock:uniform}) \citep{Ronquist:2012ea}. 
It assumes that the internal nodes are draw from uniform distributions and the fossils are only tips of the tree (so-called tip dating). 
Same as the FBD prior, the uniform prior is also conditioned on the root age and requires setting {\tt treeagepr}.

Using the molecular clock itself in this case cannot root the tree properly, we have to enable the constraint {\tt HymenFossil} defined above.
This forces the Hymenoptera with fossils to be a monophyletic group.
\begin{framed}
{\tt \noindent
prset topologypr = constraint(HymenFossil)}
\end{framed}

For the Markov chain Monte Carlo (MCMC) \citep{Metropolis:1953vj, Hastings:1970ew}, we use two independent runs and four chains (one cold and three hot) per run for 500,000 iterations and sample every 100 iterations.
\begin{framed}
{\tt \noindent
mcmcp nrun = 2 nchain = 4 ngen = 500000 samplefr = 100 \\
mcmcp filename = hym.te  printfr = 1000 diagnfr = 5000 \\
mcmc}
\end{framed}
\noindent 
The output file names are {\tt hym.te.*}. The chain states are printed to screen every 1000 iterations, and the convergence diagnostics is printed every 5000 iterations.
The {\tt mcmc} command executes the MCMC run.

While the MCMC is running, the iteration number, likelihood values, and average standard deviation of split frequencies (ASDSF) are printed to the screen.
The ASDSF should be decreasing toward 0, indicating the tree topologies sampled from different runs are getting similar and converging to the same (stationary) distribution.

To summarize the parameters and trees after the MCMC, use
\begin{framed}
{\tt \noindent
sump  \\
sumt}
\end{framed}
\noindent By default, the first 25\% samples are discarded as burnin. 
This can be adjusted according to the likelihood traces from the two runs.
The effective sample size (ESS) also helps us to judge if sufficient MCMC samples are collected.
Ideally, the ESS should be larger than 200 for all parameters.
We may need to increase the chain length or adjust the priors to improve the estimates.

The consensus tree including all fossils is highly unresolved due to the uncertainty in the placement of the fossils.
In order to display the node ages clearly, we can redraw an extant taxa tree by pruning the fossils. The output filename is changed to avoid overwriting the existing ones.
\begin{framed}
{\tt \noindent
delete fossils  \\
sumt output = hym.rf}
\end{framed}

\subsection{Node dating}

In node dating, we calibrate the internal nodes of the tree instead of assigning priors to the fossils.
The morphological characters of the fossils are not used, thus we remove them.
\begin{framed}
{\tt \noindent
delete fossils \\
exclude 24 130 168 \\
calibrate \\
\indent Tenthredinidae = offsetgamma(100,150,25) \\
\indent CepSirOruApo = truncatednormal(140,175,25) \\
prset nodeagepr = calibrated}
\end{framed}

The birth-death prior under diversified sampling \citep{Hohna:2011hd} is used for the time tree.
Compared to the FBD prior used above, there is no fossil sampling parameter in this case.
The priors for the root age, speciation and extinction rates are not changed.
\begin{framed}
{\tt \noindent
prset brlenspr = clock:birthdeath    \\
prset samplestrat = diversity        \\
prset sampleprob = 0.0001            \\
prset speciationpr = exp(10)         \\
prset extinctionpr = beta(1,1)       \\
prset treeagepr = offsetexp(300,390)}
\end{framed}
\noindent The uniform tree prior ({\tt clock:uniform}) is also applicable.

It is required to force the calibrated clade to be monophyletic and enable the constraints.
The constraint {\tt Hymenoptera} helps to root the tree properly.
\begin{framed}
{\tt \noindent
prset topologypr = constraint(Hymenoptera, Tenthredinidae, CepSirOruApo)}
\end{framed}

The output filename is changed to avoid overwriting the existing ones.
If the node dating analysis is continued after the total-evidence dating in the same MrBayes session, the starting values also need to be reset.
The other settings are kept the same as in the total-evidence dating analysis.
\begin{framed}
{\tt \noindent
mcmcp filename = hym.nd startp = reset startt = random}
\end{framed}

\subsection{Non-clock analysis}

Lastly, we run an analysis without molecular clock assumption and calibrations.
The branch lengths are measured by genetic distance (expected substitutions per site).
This is a typical analysis most people do using MrBayes.
We do not use fossils, and do not constrain the topology so that they are uniformly distributed.
\begin{framed}
{\tt \noindent
delete fossils  \\
prset brlenspr = uncons:gammadir(1,1,1,1) \\
prset topologypr = uniform  \\
mcmc filename = hym.un \\
sump  \\
sumt}
\end{framed}

The prior for branch lengths is gamma-Dirichlet(1, 1, 1, 1) \citep{Rannala:2012ke,Zhang:2012ke}, which assigns gamma(1, 1) (i.e., exp(1)) for the tree length and uniform Dirichlet for the proportion of branch lengths.
The compound Dirichlet prior was shown to help avoid overestimating the tree length \citep{Zhang:2012ke} and is now the default prior in MrBayes (since 3.2.3).

\section{Results and Discussion}

The posterior estimates of the parameters are summarized in separate files, which can be opened using a text editor.
The information is also printed to the screen.
The partition rate multipliers are in {\tt hym.*.pstat}.
The morphological ({\tt m\{1\}}) and 16S ({\tt m\{2\}}) partitions evolve at similar rate.
The 1\textsuperscript{st} and 2\textsuperscript{nd} codon positions of Ef1$\alpha$ ({\tt m\{3\}}) evolve much slower than the 3\textsuperscript{rd} codon positions ({\tt m\{4\}}).
Thus it is reasonable to take into account such significant rate variation across partitions.

\begin{figure}[h]
\center
\includegraphics[width=0.8\textwidth]{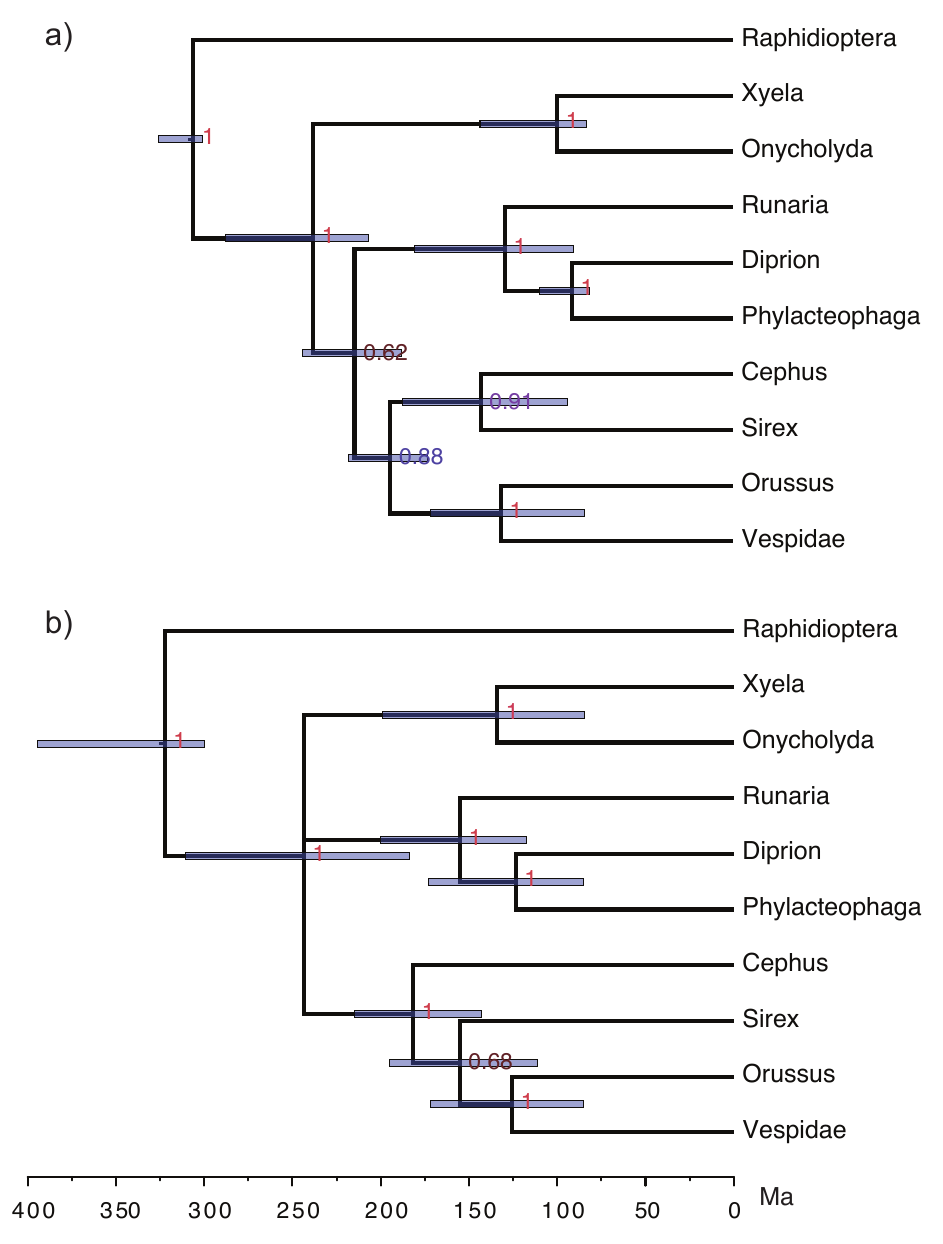}
\caption{Majority-rule consensus trees of extant taxa from a) total-evidence dating and b) node dating, under diversified sampling and IGR model. The node heights are in the unit of million years and the error bars indicate the 95\% HPD intervals. The numbers at the internal nodes are the posterior probabilities of the corresponding clades.}
\label{Fig_TimeTrees}
\end{figure}

\clearpage

The majority-rule consensus trees are summarized in {\tt hym.*.con.tre}, which can be visualized by FigTree (\url{http://tree.bio.ed.ac.uk/software/figtree}) or IcyTree (\url{https://icytree.org}).
The node ages are also in {\tt hym.*.vstat}, associated with the bipartition IDs in {\tt hym.*.parts}.
The root ID is 0 which includes all taxa, while the ID of Hymenoptera is that excludes the outgroup Raphidioptera ({\tt .*********}). 
The extant taxa trees from total-evidence dating and node dating under diversified sampling and IGR model are shown in Figure \ref{Fig_TimeTrees}.
The topologies are the same in general, except for a clade with more uncertainties. 
The mean age of Hymenoptera (250 Ma) inferred from total-evidence dating is similar to that from node dating, with relatively narrower HPD interval.

The total-evidence dating approach models the fossilization and sampling process explicitly, and incorporates different sources of information from the fossil record while accounting for the uncertainty of fossil placement.
In comparison, the node dating approach discards the fossil morphologies, and uses second interpretation of the fossil record as node calibrations.
Total-evidence dating provides an ideal platform for exploring and further improving the models used for Bayesian molecular clock dating analysis.

\begin{figure}[h]
\center
\includegraphics[width=0.7\textwidth]{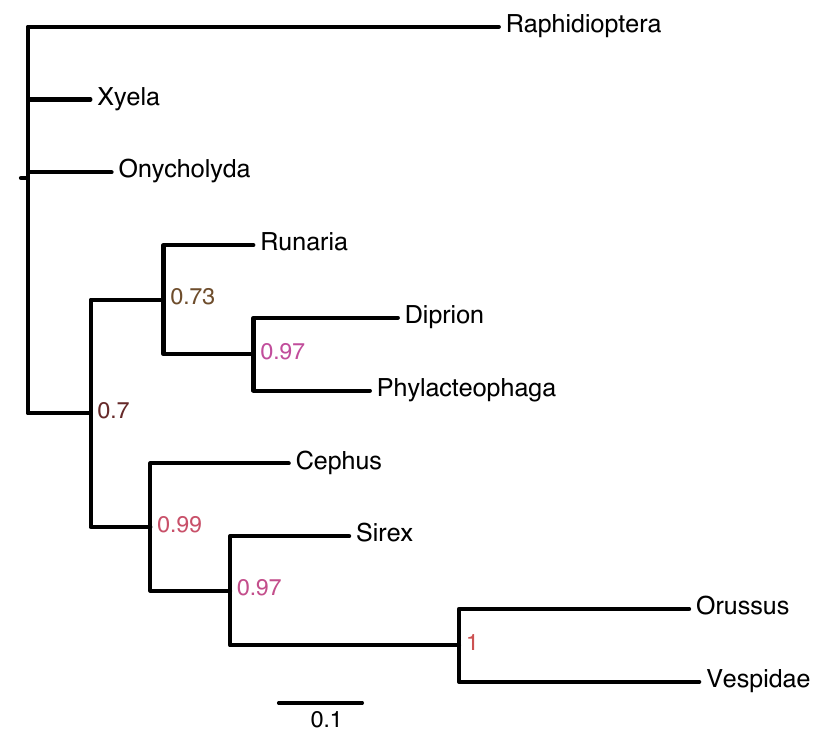}
\caption{Majority-rule consensus tree of extant taxa from a non-clock analysis under the gamma-Dirichlet prior. The branch lengths are measured by expected substitutions per site. The numbers at the internal nodes are the posterior probabilities of the corresponding clades.}
\label{Fig_NCLTree}
\end{figure}

Comparing the non-clock tree (Fig. \ref{Fig_NCLTree}) with the clock trees (Fig. \ref{Fig_TimeTrees}), it is obvious that the evolutionary rate is not constant over time.
The Xyela and Onycholyda branches evolve much slower than the Raphidioptera and Orussus/Vespidae clade, and there are indeed dramatic rate changes between adjacent branches.
Thus the IGR relaxed clock model appears more suitable than the autocorrelated TK02 model, and it is not reasonable to assume a strict clock model.
Further explorations, such as marginal likelihood estimation using stepping-stone sampling \citep{Xie:2011it}, could be carried out to compare the IGR with the TK02 model.

In conclusion, this study provides a brief overview and comparison of total-evidence dating and node dating analyses, and demonstrates the functionality of MrBayes using a dataset of Hymenoptera.
For the analyses and discussion using the full data, please see \citet{Ronquist:2012ea,Zhang:2016kf}.

\section*{Acknowledgements}

I sincerely thank Johan Nylander for valuable discussions and for organizing a workshop of MrBayes using this tutorial. 

\bibliographystyle{sysbio}
\bibliography{master_refs}

\end{document}